\documentclass[
    ,final            % use final for the camera ready runs
  ]
  {aipproc}

\layoutstyle{6x9}
\usepackage{color}
\begin{document}

\title{General properties of nonlinear mean field  Fokker-Planck equations}

\classification{05.10.Gg, 05.40.Jc, 05.45.-a, 05.90.+m }
\keywords      {Nonlinear mean field Fokker-Planck equations, Generalized entropies}

\author{P.H. Chavanis}{
  address={ Laboratoire de Physique Th\'eorique, 118 route de Narbonne
  31062 Toulouse, France} }

\begin{abstract}
Recently, several authors have tried to extend the usual concepts of
thermodynamics and kinetic theory in order to deal with distributions
that can be non-Boltzmannian. For dissipative systems described by the
canonical ensemble, this leads to the notion of nonlinear
Fokker-Planck equation (T.D. Frank, {\it Non Linear Fokker-Planck
Equations}, Springer, Berlin, 2005). In this paper, we review general
properties of nonlinear mean field Fokker-Planck equations, consider
the passage from the generalized Kramers to the generalized
Smoluchowski equation in the strong friction limit, and provide
explicit examples for Boltzmann, Tsallis and Fermi-Dirac entropies.
\end{abstract}

\maketitle

%%%%%%%%%%%%%%%%%%%%%%%%%%%%%%%%%%%%%%%%%%%%
%% MAINMATTER
%%%%%%%%%%%%%%%%%%%%%%%%%%%%%%%%%%%%%%%%%%%%

\section{1. The generalized Kramers equation}
\label{sec_gke}

\subsection{1.1. Generalized stochastic processes}
\label{sec_gsp}

Nonlinear Fokker-Planck (NFP) equations have been the subject of recent
activity \cite{frank,pp,kaniadakis,cn,pre}. Here, we consider
a generalized Kramers equation of the form \cite{cll}:
\begin{eqnarray}
\label{gsp1}
\frac{\partial f}{\partial t}+{\bf v}\cdot \frac{\partial f}{\partial {\bf r}}-\nabla\Phi\cdot  \frac{\partial f}{\partial {\bf v}}=\frac{\partial}{\partial {\bf v}}\cdot\left ( D h(f)\frac{\partial f}{\partial {\bf v}}+\xi g(f){\bf v}
\right ),
\end{eqnarray}
where $h(f)$ and $g(f)$ are positive functions. For $h(f)=1$ and
$g(f)=f$, Eq. (\ref{gsp1}) reduces to the familiar Kramers equation
where $D$ is the diffusion coefficient and $\xi$ the friction
coefficient. Usually, $\Phi({\bf r})$ is an external potential but we
can also consider the case where the potential is produced by the density
$\rho({\bf r},t)=\int f({\bf r},{\bf v},t)d{\bf v}$ according to the
relation
\begin{eqnarray}
\label{gsp2}
\Phi({\bf r},t)=\int u(|{\bf r}-{\bf r}'|)\rho({\bf r}',t)d{\bf r}',
\end{eqnarray}
where $u(|{\bf r}-{\bf r}'|)$ is a binary potential of
interaction. The nonlinear mean field Fokker-Planck equation
(\ref{gsp2}) is associated to the Ito-Langevin stochastic process
\begin{eqnarray}
\label{gsp3}
\frac{d{\bf r}}{dt}={\bf v}, \quad \frac{d{\bf v}}{dt}=-\xi(f){\bf v}-\nabla\Phi+\sqrt{2D(f)}{\bf R}(t),
\end{eqnarray}
\begin{eqnarray}
\label{gsp4}
\xi(f)=\xi\frac{g(f)}{f}, \quad D(f)=\frac{D}{f}\int_{0}^{f}h(x)dx,
\end{eqnarray} 
where ${\bf R}(t)$ is a white noise satisfying $\langle {\bf R}(t)\rangle={\bf 0}$ and $\langle R_{i}(t)R_{j}(t')\rangle =\delta_{ij}\delta(t-t')$ where $i=1,...,d$ label the coordinates of space.

\subsection{1.2. The H-theorem}
\label{sec_ht}

We introduce the energy
\begin{eqnarray}
\label{ht1}
E=\frac{1}{2}\int f v^{2}\, d{\bf r}d{\bf v}+\frac{1}{2}\int \rho \Phi\, d{\bf r}=K+W,
\end{eqnarray} 
where $K$ is the kinetic energy and $W$ is the potential energy. For an external potential, we have $W=\int \rho \Phi d{\bf r}$. We define the temperature by
\begin{eqnarray}
\label{ht2}
T=\frac{D}{\xi}.
\end{eqnarray} 
Therefore, the Einstein relation is preserved in this generalized thermodynamical framework. We also set $\beta=1/T$. We introduce the generalized
entropic functional
\begin{eqnarray}
\label{ht3}
S=-\int C(f)\, d{\bf r}d{\bf v},
\end{eqnarray}
where $C(f)$ is a convex function ($C''>0$) satisfying \cite{cll}:
\begin{eqnarray}
\label{ht4}
C''(f)=\frac{h(f)}{g(f)}.
\end{eqnarray}
Since the temperature is fixed (canonical description), the relevant thermodynamical potential is the generalized free energy
\begin{eqnarray}
\label{ht5}
F=E-TS.
\end{eqnarray}
The definition of the free energy (Legendre transform) is preserved in
this generalized thermodynamical framework. A straightforward
calculation then shows that \cite{cll}:
\begin{eqnarray}
\label{ht6}
\dot F=
-\int \frac{1}{\xi g(f)}\left (Dh(f)\frac{\partial f}{\partial {\bf v}}+\xi g(f){\bf v}\right )^{2}d{\bf r}d{\bf v}.
\end{eqnarray}
Therefore, $\dot F\le 0$ (provided that $\xi>0$). This forms an
$H$-theorem in the canonical ensemble. The free energy $F$ plays the
role of a Lyapunov functional. Note that the NFP equation (\ref{gsp1}) can be written
\begin{eqnarray}
\label{ht7}
\frac{\partial f}{\partial t}+{\bf v}\cdot \frac{\partial f}{\partial {\bf r}}-\nabla\Phi\cdot  \frac{\partial f}{\partial {\bf v}}=\frac{\partial}{\partial {\bf v}}\cdot \left\lbrack \xi g(f)\frac{\partial}{\partial {\bf v}}\left (\frac{\delta F}{\delta f}\right )\right\rbrack.
\end{eqnarray}

\subsection{1.3. Stationary solutions}
\label{sec_ss}

The steady states of Eq. (\ref{gsp1}) must satisfy $\dot F=0$ leading to a vanishing current
\begin{eqnarray}
\label{ss1}
{\bf J}\equiv Dh(f)\frac{\partial f}{\partial {\bf v}}+\xi g(f){\bf v}={\bf 0}.
\end{eqnarray}
Using Eqs. (\ref{ht2}) and (\ref{ht4}), we get
\begin{eqnarray}
\label{ss2}
C''(f)\frac{\partial f}{\partial {\bf v}}+\beta {\bf v}={\bf 0},
\end{eqnarray}
which can be integrated into
\begin{eqnarray}
\label{ss3}
C'(f)=-\beta \left\lbrack \frac{v^{2}}{2}+\lambda({\bf r})\right\rbrack,
\end{eqnarray}
where $\lambda({\bf r})$ is a function of the position. Since
$\partial f/\partial t=0$ and ${\bf J}={\bf 0}$, the advective
(Vlasov) term in Eq. (\ref{gsp1}) must also vanish leading to the
condition
\begin{eqnarray}
\label{ss4}
{\bf v}\cdot \frac{\partial f}{\partial {\bf r}}-\nabla\Phi\cdot  \frac{\partial f}{\partial {\bf v}}=0.
\end{eqnarray}
Using 
\begin{eqnarray}
\label{ss5}
C''(f)\frac{\partial f}{\partial {\bf r}}=-\beta\nabla \lambda, \quad C''(f)\frac{\partial f}{\partial {\bf v}}=-\beta {\bf v},
\end{eqnarray}
we obtain $(\nabla\lambda-\nabla\Phi)\cdot {\bf v}=0$
which must be true for all ${\bf v}$. This yields $\nabla\lambda-\nabla\Phi={\bf 0}$, so that
\begin{eqnarray}
\label{ss6}
\lambda({\bf r})=\Phi({\bf r})+\alpha/\beta,
\end{eqnarray}
where $\alpha$ is a constant. Therefore, the stationary solutions of  Eq. (\ref{gsp1}) are determined by the relation \cite{pre}:
\begin{eqnarray}
\label{ss7}
C'(f)=-\beta\epsilon-\alpha,
\end{eqnarray}
where $\epsilon={v^{2}}/{2}+\Phi({\bf r})$ is the individual energy. 
Since $C$ is convex, this equation can be reversed to give 
\begin{eqnarray}
\label{ss8}
f=F(\beta \epsilon+\alpha),
\end{eqnarray}
where $F(x)=(C')^{-1}(-x)$ is a decreasing function. Thus $f=f(\epsilon)$ is a decreasing function of the energy. We have $f'(\epsilon)=-\beta/C''(f)\le 0$. 

\subsection{1.4. Minimum of free energy}
\label{sec_min}

The critical points of free energy at fixed mass are determined by the variational problem
\begin{eqnarray}
\label{min1}
\delta F+T\alpha\delta M=0,
\end{eqnarray}
where $\alpha$ is a Lagrange multiplier. We can easily establish that
\begin{eqnarray}
\label{min2}
\delta E=\int \left (\frac{v^{2}}{2}+\Phi \right )\ \delta f  \, d{\bf r}d{\bf v},\quad \delta S=-\int C'(f)\delta f \, d{\bf r} d{\bf v}.
\end{eqnarray}
Therefore, the variational principle (\ref{min1}) gives \cite{pre}:
\begin{eqnarray}
\label{min3}
C'(f)=-\beta\epsilon-\alpha,
\end{eqnarray}
equivalent to Eq. (\ref{ss7}).  Therefore, a stationary solution of
the GK equation (\ref{gsp1}) is a critical point of free energy $F[f]$ at
fixed mass $M$. Furthermore, it is shown in Ref. \cite{frank,pre} that
it is linearly dynamically stable if and only if it is a {\it minimum}
(at least local) of $F$ at fixed mass. Note that when $\Phi$ is an
external potential, we have $\delta^{2}F=-T\delta^{2}S=(1/2)T\int
C''(f){(\delta f)^{2}}d{\bf r}d{\bf v}\ge 0$ so that a critical point
of $F$ is always a minimum.

\section{2. The generalized Smoluchowski equation}
\label{sec_gs}

We restrict ourselves to the case of a constant friction so that $g(f)=f$ and
$h(f)=fC''(f)$. The generalized Kramers equation (\ref{gsp1}) then becomes
\begin{eqnarray}
\label{gs1}
\frac{\partial f}{\partial t}+{\bf v}\cdot \frac{\partial f}{\partial {\bf r}}-\nabla\Phi\cdot  \frac{\partial f}{\partial {\bf v}}=\frac{\partial}{\partial {\bf v}}\cdot\left \lbrack \xi \left (TfC''(f)\frac{\partial f}{\partial {\bf v}}+f{\bf v}
\right )\right\rbrack.
\end{eqnarray}
In that case, $\xi(f)=\xi$ and $D(f)=Df \lbrack C(f)/f\rbrack'$. Let us derive the hydrodynamic moments of this equation \cite{pre}. Defining
the density and the local velocity by
\begin{eqnarray}
\label{gs2} \rho=\int f\,d{\bf v}, \qquad \rho{\bf u}=\int f{\bf
v}\,d{\bf v},
\end{eqnarray}
and integrating Eq.~(\ref{gs1}) on velocity, we get the continuity equation
\begin{eqnarray}
\label{gs3} {\partial\rho\over\partial t}+\nabla\cdot (\rho{\bf u})=0.
\end{eqnarray}
Next, multiplying Eq.~(\ref{gs1}) by ${\bf v}$, integrating on the 
velocity and using the continuity equation (\ref{gs3}), we obtain the momentum equation
\begin{eqnarray}
\label{gs4} \rho \left ({\partial u_{i}\over\partial t}+u_{j}{\partial u_{i}\over\partial x_{j}}\right )= -{\partial
P_{ij}\over\partial x_{j}}-\rho{\partial\Phi\over\partial
x_{i}}-\xi\rho u_i,
\end{eqnarray}
where we have defined the pressure tensor
\begin{eqnarray}
\label{gs5}P_{ij}=\int fw_{i}w_{j}\,d{\bf v},
\end{eqnarray}
where ${\bf w}={\bf v}-{\bf u}({\bf r},t)$ is the relative
velocity. We now derive the generalized Smoluchowski (GS) equation from the
generalized Kramers (GK) equation in the strong friction limit (see
\cite{banach}, Sec. 9).  For $\xi\rightarrow +\infty$ with fixed $T$,
the term in parenthesis in Eq. (\ref{gs1}) must vanish at leading order
\begin{eqnarray}
\label{gs6}
TfC''(f)\frac{\partial f}{\partial {\bf v}}+f{\bf v}\simeq {\bf 0}.
\end{eqnarray}
Then, we find that the out-of-equilibrium
distribution function $f_0({\bf r},{\bf v},t)$ is determined by
\begin{eqnarray}
\label{gs7} C'(f_0)=-\beta\left\lbrack \frac{v^{2}}{2}+\lambda({\bf
r},t)\right\rbrack+O(\xi^{-1}),
\end{eqnarray}
where $\lambda({\bf r},t)$ is a constant of integration that is
determined by the density according to
\begin{eqnarray}
\label{gs8} \rho({\bf r},t)=\int f_{0}d{\bf v}=\rho[\lambda({\bf
r},t)].
\end{eqnarray}
Note that the distribution function $f_0$ is {\it isotropic} so that
the velocity ${\bf u}({\bf r},t)=O(\xi^{-1})$ and the pressure
tensor  $P_{ij}=p\delta_{ij}+O(\xi^{-1})$ where $p$ is given by
\begin{eqnarray}
\label{gs9} p({\bf r},t)=\frac{1}{d}\int f_{0}v^{2}d{\bf
v}=p[\lambda({\bf r},t)].
\end{eqnarray}
Eliminating $\lambda({\bf r},t)$ between Eqs. (\ref{gs8}) and
(\ref{gs9}), we find that the fluid is barotropic in the sense that
$p=p(\rho)$, where the equation of state is entirely determined by the
generalized entropy $C(f)$. Now, considering the momentum equation
(\ref{gs4}) in the limit $\xi\rightarrow +\infty$, we find that
\begin{eqnarray}
\label{gs10} \rho{\bf u}=-\frac{1}{\xi}(\nabla
p+\rho\nabla\Phi)+O(\xi^{-2}).
\end{eqnarray}
Inserting this relation in the continuity equation  (\ref{gs3}), we obtain
the generalized Smoluchowski equation \cite{pre}:
\begin{eqnarray}
\label{gs11}\frac{\partial\rho}{\partial t}=\nabla\cdot
\left\lbrack \frac{1}{\xi}(\nabla p+\rho\nabla\Phi)\right\rbrack.
\end{eqnarray}
This equation can also be obtained from a Chapman-Enskog expansion in
powers of $1/\xi$ \cite{cll}. It monotonically decreases the free energy \cite{pre}:
\begin{eqnarray}
\label{gs12} F[\rho]=\int \rho\int^{\rho}{p(\rho')\over
\rho'^{2}} \,d \rho'd{\bf r}+{1\over 2}\int\rho\Phi d{\bf r},
\end{eqnarray}
which can be deduced from the free energy (\ref{ht5}) by using Eq. (\ref{gs7})
to express $F[f]$ as a functional $F[\rho]\equiv F[f_0]$ of the density
\cite{cll,banach}.  A direct calculation leads to the $H$-theorem
\begin{eqnarray}
\label{gs13} \dot F=
-\int \frac{\xi}{\rho}(\nabla p+\rho\nabla \Phi)^{2}d{\bf r}\le 0.
\end{eqnarray}
Moreover the stationary solutions of the generalized Smoluchowski
equation (\ref{gs11}) are critical points of the free energy $F[\rho]$
at fixed mass, satisfying $\delta F-\alpha\delta M=0$ where $\alpha$
is a Lagrange multiplier. This yields
$\int^{\rho}p'(\rho')/\rho'\, d\rho'=-\Phi$ leading to the condition of hydrostatic balance
\begin{eqnarray}
\label{gs14}\nabla p+\rho\nabla\Phi={\bf 0}.
\end{eqnarray}
After integration, we get $\rho=\rho(\Phi)$ with $\rho'(\Phi)\le
0$. This result can also be obtained by integrating $f=f(\epsilon)$ on
the velocity. Finally, a steady state of the GS
equation (\ref{gs11}) is linearly dynamically stable iff it is a
(local) minimum of $F[\rho]$ at fixed mass \cite{frank,pre}.

\section{3. Explicit examples}
\label{sec_ex}

\subsection{3.1. Isothermal systems: Boltzmann entropy}
\label{sec_iso}

If we consider the Boltzmann entropy
\begin{equation}
\label{iso1} S_{B}[f]=-\int f\ln f d{\bf r}d{\bf v},
\end{equation}
we get the ordinary Kramers equation
\begin{eqnarray}
\label{iso2}
\frac{\partial f}{\partial t}+{\bf v}\cdot \frac{\partial f}{\partial {\bf r}}-\nabla\Phi\cdot  \frac{\partial f}{\partial {\bf v}}=\frac{\partial}{\partial {\bf v}}\cdot\left \lbrack \xi \left (T\frac{\partial f}{\partial {\bf v}}+f{\bf v}
\right )\right\rbrack.
\end{eqnarray}
The stationary solution is the Boltzmann distribution
\begin{equation}
\label{iso3} f=A e^{-\beta\epsilon},
\end{equation}
where $A$ is determined by the conservation of mass. The equation of
state is the isothermal one
\begin{equation}
\label{iso4} p=\rho T.
\end{equation}
In the strong friction limit, we recover the ordinary Smoluchowski equation
\begin{eqnarray}
\label{iso5} {\partial\rho\over\partial t}=\nabla \cdot \biggl\lbrack
{1\over\xi}(T\nabla \rho+\rho\nabla\Phi)\biggr\rbrack.
\end{eqnarray}
The free energy is the Boltzmann free energy in physical space
\begin{eqnarray}
{F}[\rho]=T\int \rho \ln\rho\ d{\bf r}+{1\over
  2}\int\rho\Phi \ d{\bf r}.
\label{iso6}
\end{eqnarray}
The stationary solution is  the Boltzmann distribution in physical space
\begin{equation}
\label{iso7} \rho=A' e^{-\beta\Phi},
\end{equation}
where $A'=(2\pi/\beta)^{d/2}A$.

\subsection{3.2. Polytropes: Tsallis entropy}
\label{sec_poly}

If we consider the Tsallis $q$-entropy
\begin{equation}
\label{poly1} S_{q}[f]=-{1\over q-1}\int (f^{q}-f)  d{\bf r} d{\bf
v},
\end{equation}
we obtain the polytropic Kramers equation
\begin{eqnarray}
\label{poly2}
\frac{\partial f}{\partial t}+{\bf v}\cdot \frac{\partial f}{\partial {\bf r}}-\nabla\Phi\cdot  \frac{\partial f}{\partial {\bf v}}=\frac{\partial}{\partial {\bf v}}\cdot\left \lbrack \xi \left (T\frac{\partial f^{q}}{\partial {\bf v}}+f{\bf v}
\right )\right\rbrack.\nonumber\\
\end{eqnarray}
The stationary solution is the Tsallis (or polytropic)  distribution 
\begin{equation}
\label{poly3} f=\biggl\lbrack \mu-{(q-1)\beta\over
q}\epsilon\biggr\rbrack_{+}^{1\over q-1},
\end{equation}
where $\mu$ is determined by the conservation of mass.  The index $n$
of the polytrope is
\begin{equation}
\label{poly4} n=\frac{d}{2}+\frac{1}{q-1}.
\end{equation}
Isothermal distribution functions are recovered in the limit
$q\rightarrow 1$ (i.e. $n\rightarrow +\infty$). We shall consider
$q>0$ so that $C$ is convex. For $q>1$, i.e. $n>d/2$, the distribution
has a compact support (case 1) since $f$ is defined only for
$\epsilon\le
\epsilon_{m}=q\mu/\lbrack (q-1)\beta\rbrack$. For $\epsilon\ge \epsilon_{m}$, we set
$f=0$. For $n=d/2$, $f$ is the Heaviside function. For $q<1$, the
distribution is defined for all energies (case 2). For large
velocities, it behaves like $f\sim v^{2n-d}$. Therefore, the density
and the pressure are finite only for $n<-1$, i.e. $d/(d+2)<q<1$. This
fixes the range of allowed parameters. The equation of state is that
of a polytrope \cite{cs,csa}
\begin{equation}
\label{poly5} p=K\rho^{\gamma}, \qquad \gamma=1+{1\over n}.
\end{equation}
For $n>d/2$ (case 1) the polytropic constant is
\begin{equation}
\label{poly6} K=\frac{1}{n+1}\left\lbrack A S_{d}
2^{\frac{d}{2}-1}\frac{\Gamma\left (d/2\right )\Gamma\left
(1-d/2+n\right )}{\Gamma(1+n)}\right \rbrack^{-1/n},
\end{equation}
and for $n<-1$ (case 2), we have
\begin{equation}
\label{poly7} K=-\frac{1}{n+1}\left\lbrack A S_{d}
2^{\frac{d}{2}-1}\frac{\Gamma\left (d/2\right )\Gamma\left (-n\right
)}{\Gamma(d/2-n)}\right \rbrack^{-1/n},
\end{equation}
where $A=\lbrack\beta|q-1|/q\rbrack^{1/(q-1)}$. In the strong friction
limit, we get the polytropic Smoluchowski equation
\begin{eqnarray}
\label{poly8} {\partial\rho\over\partial t}=\nabla \cdot \biggl\lbrack
{1\over\xi}(K\nabla \rho^{\gamma}+\rho\nabla\Phi)\biggr\rbrack.
\end{eqnarray}
The free energy is the Tsallis free energy in physical space
\begin{eqnarray}
{F}[\rho]={K\over\gamma -1}\int (\rho^{\gamma}-\rho) \ d{\bf
r}+{1\over
  2}\int\rho\Phi \ d{\bf r}.
\label{poly9}
\end{eqnarray}
The stationary solution is the Tsallis distribution in physical space
\begin{equation}
\label{poly10} \rho=\biggl\lbrack \lambda-{\gamma-1\over
K\gamma}\Phi\biggr\rbrack_{+}^{1\over\gamma-1}.
\end{equation}
We note that a polytropic distribution with index $q$ in phase space
yields a polytropic distribution with index $\gamma=1+2(q-1)/\lbrack
2+d(q-1)\rbrack$ in physical space. In this sense, Tsallis
distributions are stable laws. By comparing Eq. (\ref{poly3}) with
Eq. (\ref{poly10}) or Eqs. (\ref{ht5}) and (\ref{poly1}) with
Eq. (\ref{poly9}) we note that $K$ plays the same role in physical
space as the temperature $T=1/\beta$ in phase space. It is sometimes
called a ``polytropic temperature'' \cite{csa}.

\subsection{3.3. Fermions: Fermi-Dirac entropy}
\label{sec_fermi}

If we consider the Fermi-Dirac entropy
\begin{equation}
\label{fermi1} S_{FD}[f]=-\eta_{0}\int \biggl\lbrace {f\over \eta_{0}}\ln
{f\over \eta_{0}}+\biggl (1-{f\over \eta_{0}}\biggr )\ln \biggl
(1-{f\over \eta_{0}}\biggr ) \biggr\rbrace d{\bf r} d{\bf v},
\end{equation}
we obtain the fermionic Kramers equation
\begin{eqnarray}
\label{fermi2}
\frac{\partial f}{\partial t}+{\bf v}\cdot \frac{\partial f}{\partial {\bf r}}-\nabla\Phi\cdot  \frac{\partial f}{\partial {\bf v}}
=\frac{\partial}{\partial {\bf v}}\cdot\left \lbrack \xi \left (-T\eta_{0}\frac{\partial}{\partial {\bf v}}\ln\left (1-\frac{f}{\eta_{0}}\right )+f{\bf v}
\right )\right\rbrack.
\end{eqnarray}
The stationary solution is the Fermi-Dirac distribution function
\begin{equation}
\label{fermi3} f={\eta_{0}\over 1+\lambda e^{\beta\epsilon}},
\end{equation}
where $\lambda>0$ is determined by the conservation of mass. The
Fermi-Dirac distribution function (\ref{fermi3}) satisfies the
constraint $f\le \eta_{0}$ which is related to the Pauli exclusion
principle in quantum mechanics. The isothermal distribution function
(\ref{iso3}) is recovered in the non-degenerate limit $f\ll \eta_{0}$
(valid at high temperatures). On the other hand in the completely
degenerate limit (valid at low temperatures) the distribution is a
step function corresponding to a polytrope of index $n=d/2$. The
distribution in physical space associated with the Fermi-Dirac
statistics is
\begin{equation}
\label{fermi4} \rho={\eta_{0}S_{d}2^{{d\over
2}-1}\over\beta^{d/2}}I_{{d\over 2}-1}(\lambda e^{\beta\Phi}),
\end{equation}
where $I_{n}$ is the Fermi integral
\begin{equation}
\label{fermi5} I_{n}(t)=\int_{0}^{+\infty}{x^{n}\over 1+te^{x}}dx.
\end{equation}
The quantum equation of state  for fermions is given in parametric form by
\begin{equation}
\label{fermi6} \rho={\eta_{0}S_{d}2^{{d\over
2}-1}\over\beta^{d/2}}I_{{d\over 2}-1}(t),\qquad
p={\eta_{0}S_{d}2^{{d\over 2}}\over d\beta^{{d\over 2}+1}}I_{{d\over
2}}(t).
\end{equation}
At high temperatures we recover the classical isothermal law $p=\rho
T$ and at low temperatures we get a polytropic equation of state
$p=K\rho^{\gamma}$ with $\gamma=\frac{d+2}{2}$ and $K=\frac{1}{d+2}(\frac{d}{\eta_{0}S_{d}})^{2/d}$.



\bibliographystyle{aipproc}   % if natbib is available

\bibliography{sample}

\IfFileExists{\jobname.bbl}{}
 {\typeout{}
  \typeout{******************************************}
  \typeout{** Please run "bibtex \jobname" to optain}
  \typeout{** the bibliography and then re-run LaTeX}
  \typeout{** twice to fix the references!}
  \typeout{******************************************}
  \typeout{}
 }

%%%%%%%%%%%%%%%%%%%%%%%%%%%%%%%%%%%%%%%%%%%
%% The following lines show an example how to produce a bibliography
%% without the help of the BibTeX program. This could be used instead
%% of the above.
%%%%%%%%%%%%%%%%%%%%%%%%%%%%%%%%%%%%%%%%%%%

\end{document}